# Robust Analysis of Stock Price Time Series Using CNN and LSTM-Based Deep Learning Models

Sidra Mehtab
Department of Data Science
Praxis Business School
Kolkata, INDIA
email: smehtab@acm.org

Jaydip Sen
Department of Data Science
Praxis Business School
Kolkata, INDIA
email: jaydip.sen@acm.org

Subhasis Dasgupta
Department of Data Science
Praxis Business School
Kolkata, INDIA
email: subhasis@praxis.ac.in

*Abstract*— Prediction of stock price and stock price movement patterns has always been a critical area of research. While the well-known efficient market hypothesis rules out any possibility of accurate prediction of stock prices, there are formal propositions in the literature demonstrating accurate modeling of the predictive systems can enable us to predict stock prices with a very high level of accuracy. In this paper, we present a suite of deep learning-based regression models that yields a very high level of accuracy in stock price prediction. To build our predictive models, we use the historical stock price data of a well-known company listed in the National Stock Exchange (NSE) of India during the period December 31, 2012 to January 9, 2015. The stock prices are recorded at five minutes interval of time during each working day in a week. Using these extremely granular stock price data, we build four *convolutional neural network* (CNN) and five *long- and short-term memory* (LSTM)-based deep learning models for accurate forecasting of the future stock prices. We provide detail results on the forecasting accuracies of all our proposed models based on their execution time and their *root mean square error* (RMSE) values.

I. INTRODUCTION

Analysis of financial time series and prediction of future stock prices and future stock price movement patterns have been an active area of research over a considerable period of time. While there are people who believe in the well-known efficient *market hypothesis* and claim that it is impossible to forecast stock prices accurately, propositions exist in the literature that demonstrate that it is possible to predict the values of stock prices with a very high level of accuracy using optimally designed and fine-tuned models. The latter have focused on the choice of variables, appropriate functional forms, and techniques of forecasting. Time series decomposition of stock price data is also a popular approach for stock price forecasting [1-2]. The use of machine learning and deep learning-based approaches are also adopted in some propositions [3]. An approach based on text mining and natural language processing in analyzing the sentiments in the social media, and utilizing that information in building a non-linear predictive model for accurate stock price prediction has been proposed in the literature [4]. Deployment of a suite of *convolutional neural networks* (CNN) for achieving a very high level of accuracy in the forecasting of the stock price has also been proposed [5].

Numerous contributions exist in the literature on technical analysis of stock prices that attempt to identify patterns in stock price movement so that profitable decisions can be made on stock market investments. A gamut of indicators has been proposed in the literature for characterizing the behavior of stock price movement. Some of the important indicators like *moving average convergence divergence* (MACD) and *meta sine wave* etc., provide the potential investors with visual representations enabling them to make wise decisions in investments in the stock market.

In this work, we present a suite of regression models for forecasting of future stock prices of a well-known company listed in the National Stock Exchange (NSE) of India. The proposition includes three regression models that are built on *convolutional neural networks* (CNNs) and four predictive models based on *long-and-short-term memory* (LSTM) networks. The models differ in their architectural designs and input data structures.

The contributions of the paper are three-fold. First, unlike most of the existing propositions in the literature, the models presented here are designed to handle extremely granular stock price data collected at 5 min interval of time. Second, our propositions include deep learning models that provide a very high level of accuracy in stock price forecasting. The most accurate model in this work yields a value of 0.00625 as a ratio of root mean square error (RMSE) to the mean value of the target variable in the test data set. Third, the proposed models are very fast in execution. The fastest model took to 83.42s to execute over a training dataset consisting of 19500 records and a test dataset with 20500 records on the target hardware platform.

The rest of the paper is organized as follows. In Section II, we very briefly discuss some related work on forecasting of stock prices. Section III presents a detailed discussion on the stock price data that we have used, and the research methodology we followed in this work. Detailed experimental results on the performance of our proposed suite of models are presented in Section IV. Finally, we conclude the paper in Section V.

II. RELATED WORK

Forecasting of future stock prices and stock price movement patterns have been an active area of research over a long period of time. The propositions are quite diverse in their methodology and algorithms used by them. However, most of the approaches can broadly categorized into three groups. The propositions belonging to the first category follow *ordinary least square* (OLS) regression or some form of its variants on cross-sectional data [6-8]. Unfortunately, models using this approach fail to achieve high accuracy in forecasting. This is because of the fact that the assumptions made by the models are, most often, not satisfied by the real-world data. The propositions of the second category use time series models using econometric approaches like *autoregressive integrated moving average* (ARIMA), *vector autoregression* (VAR), and *autoregressive distributed lag* (ARDL) [9-11]. Although, these models perform well on financial time series data with dominant trend and seasonality components, their performance on volatile data exhibiting high degree of randomness has been suboptimal. The models in the third category are *learning-based systems* that use various algorithm of machine learning, deep learning and natural language processing on structured and unstructured data including textual information in the social web [12-14]. The predictive models belonging to the third category are found to be most effective in handling volatile and extremely granular financial time series data.

The most common drawback of the majority of the existing approaches for stock price prediction is their inability to effectively handle randomness in financial time series data. Our current work attempts to address this problem by exploiting the high learning abilities of *convolutional neural networks* (CNNs) and *long-and-short-term memory* (LSTM) networks.

### III. METHODOLOGY

In this work, our main goal is to build a framework of predictive models for accurately forecasting the *open* value of the stock price of the company *Bharat Forge* listed in the NSE, India. We use stock price data for the company *Bharat Forge* for the period December 31, 2012 (which was a Monday) to January 9, 2015 (which was a Friday). During this period of time, the stock price data have been captured at an interval of 5 minutes by the *Metastock* tool [15]. For building the models, we use the stock price data for the period: December 31, 2012 to December 30, 2013. The models are tested on the historical stock price data from December 31, 2013 to January 9, 2015. The entire dataset has also been organized in a form of a weekly sequence of data from Monday to Friday. The training and the test datasets consist of 19500 and 20500 records respectively. Each record consists of the following attributes: (i) *date*, (ii) *time* slot, (iii) *open*, (iv) *high*, (v) *low*, (vi) *close*, and (vii) *volume*.

We build seven deep learning-based regression models for forecasting stock prices. In the univariate models, the variable *open* is used as the target variable, and the future *open* values are predicted using its past values. In the multivariate models, while *open* is still used as the response variable, other variables, e.g., *high*, *low*, *close*, and *volume* are used as the predictors. For testing the models, an approach called *multi-step forecasting with walk-forward validation* approach is followed [8]. Following this method, the models are built using the records in the training dataset, and then the *open* values of the *Bharat Forge* stock for the next are forecasted. At the completion of one week, the actual *open* values of that week are included in the training dataset, and then the forecasts for the *open* values for the next week are made. At each round, forecasting of the *open* values for the five days in the upcoming week is made.

We have demonstrated the efficacy and efficiency of *convolutional neural networks* (CNNs) in time series analysis and forecasting in one of our recently published work [5]. In the current work, in addition to exploiting the power of CNN, we propose the application of *long-and-short-term memory* (LSTM) networks in analysis and forecasting on a complex multivariate time series.

CNNs have two processing layers that are responsible for carrying out the major computations [5]. While the convolutional layers are responsible for identifying the important features from the input data, the pooling or sub-sampling layers summarize those features and extract the most prominent among them in a neighborhood. The final pooling layer feeds its output into one or more dense layers, and the classification or regression tasks are performed.

LSTM is a variant of deep neural network that has the capability to read and interpret sequential data like text or time series [14]. LSTM networks have the ability to maintain their state information using memory cells and gates. The gates enable these networks to reject irrelevant information of the past, remember important information in the current state, and capture the input to the system at the current instant of time, in order to produce the output as the forecast for the next time instant. The state vector in the LSTM memory cell carries out aggregation of the old information received from the *forget gates*, and the most recent information received from the *input gates*. Finally, the *output gates* produce the output from the network at the current slot. This output can be considered as the forecasted value computed by the model for the next time [14].

We propose seven different deep learning-based predictive models in this work. The models vary in their design, architecture, and input data shapes. Among these models, three are based on CNN architecture and the remaining four are built on LSTM networks. The models are: (i) CNN model with univariate input data of the previous one week, (ii) CNN model with univariate input data of the previous two weeks, (iii) CNN model with multivariate input data of the previous two weeks, (iv) LSTM model with univariate input data of previous one week, (v) LSTM model with univariate input data of previous two weeks, (vi) Encoder-decoder LSTM model with univariate data of previous two weeks, (vii) Encoder-decoder LSTM model with multivariate input data of previous two weeks.

In the following, a brief outline on the architectural details of the above models is presented. The details of these models can be found in [5, 14].

The first model is built on a CNN architecture that takes the previous one week's data as the input and makes forecasting of the next week's *open* values in a *multi-step walk-forward* manner. The input data shape is (5,1) indicating that only one attribute (i.e., *open*) of the stock price time series is used for the five days in the previous week. The model deploys one convolutional layer and one max-pooling layer. The output of the max-pooling layer is reshaped into a flat one-dimensional vector, and then the flattened vectors is allowed to pass into a fully-connected layer. Finally, the output layer predicts the *open* values for the next five days. The model is trained using 20 epochs and a batch size of 4 using a *rectified linear unit* (ReLU) as the activation function in the convolutional and the max-pooling layer, and ADAM as the optimizer. The architecture of the model is presented in Fig. 1. We refer to this model as CNN#1 model.

The second model is also a CNN model that uses the previous two weeks' *open* values as its input, and forecasts the next week's *open* values in a univariate manner. The architecture of the model is identical to that of CNN#1 model, except for the input data shape, which is (10, 1) for this model. We refer to this model as the CNN#2 model.

The third model is a multivariate CNN model that uses the previous two week's data as its input. Each of the five variables, i.e., *open*, *high*, *low*, *close*, and *volume* is used as a separate channel in a CNN. Two convolutional layers with 32 filter maps with a kernel size of 3 are deployed in this model. A max-pooling layer follows the second convolutional layer. Another convolutional layer followed by a second max-pooling layer process the data. One fully connected layer with 100 nodes processes the data after the second max-pooling layer, before the output layer produces the forecast of the next five *open* values. Fig. 2 presents the architecture of the model, which we call as CNN#3 model.

LSTM layer consisting of 200 nodes. The output of the LSTM layer propagates through a dense layer that has 200 nodes at its input and 100 nodes at the output. Finally, the forecasted values are produced by the output layer that is connected to the fully-connected layer. The output layer has 100 nodes at its input and 5 nodes at the output. Fig.3 shows the architecture of the model, which we refer to as LSTM#1.

The fifth model and the second model in the LSTM suite is also a univariate model with the previous two weeks' *open* values as the input. The architectural design and all parameters of the models are all identical to those of the LSTM#1 model, except for the input data shape which is (10,1). We call this model as LSTM#2.

The sixth model and the third one in the LSTM suite is an encoder-decoder univariate model with the past two weeks' data as the input [14]. We call this model as LSTM#3, which is shown in Fig. 4.

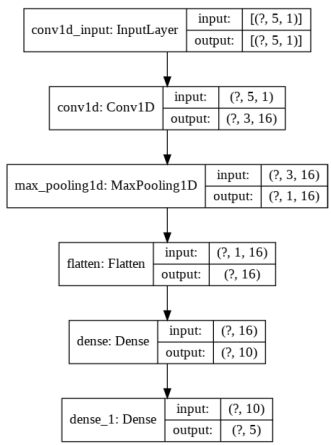

Fig. 1. The architecture of univariate CNN model (CNN#1) with prior one week's data as the input

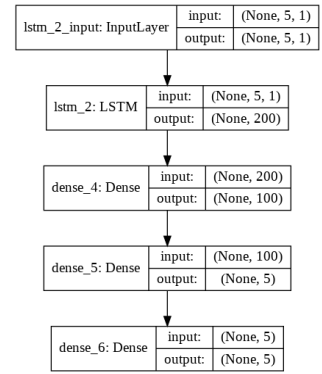

Fig. 3. The architecture of the univariate LSTM model (LSTM #1) with the previous one week's data as the input

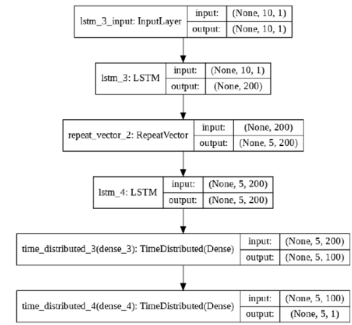

Fig. 4. The architecture of univariate encoder-decoder LSTM model (LSTM#3) with prior two weeks' data as the input

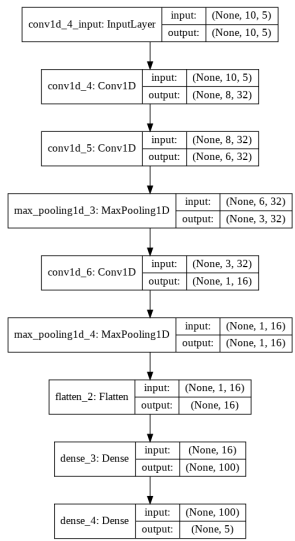

Fig. 2. The architecture of multivariate CNN model (CNN# 3) with prior two weeks' data as the input

The fourth model and the first of the LSTM suite is a univariate LSTM model that uses the previous one week's *open* values to forecast the *open* values in the next week. The shape of the input data to the model is (5,1) as in the case of the CNN#1 model. The input data is passed through an

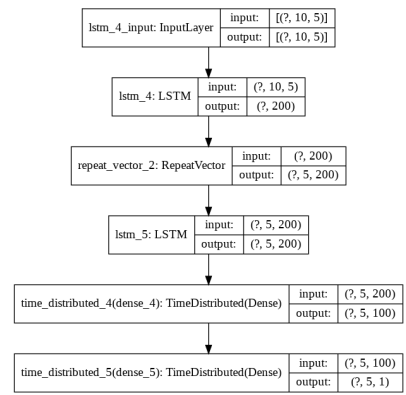

Fig. 5. The architecture of the multivariate encoder-decoder LSTM with prior two weeks' data as input (LSTM #4)

The seventh and the fourth model in the LSTM suite is an encoder-decoder LSTM model that uses a multivariate time series data as its input. The model receives the previous two weeks' stock price data with all the variables. i.e., *open*, *high*, *low*, *close*, and *volume*, as the inputs. Based on this multivariate input, the model forecasts the *open* values of the stock for the five days in the next week. Fig. 5 shows the architecture of this model. We call this model LSTM #4.

## IV. Performance Results

In this Section, we present the performance results of the deep learning models. Each model is tested over 10 rounds, and the performance of the model is noted in terms of the overall RMSE, the RMSE values for the individual days of a week (i.e., Monday – Friday), the time needed to execute one round of the model, and the ratio of the RMSE to the mean of the actual *open* value in the test dataset. It may be noted that the number of records in the training and the test dataset were 19500 and 20250 respectively. The mean *open* value in the test dataset is 628.53. The model has been executed on a computing machine consisting of an Intel i7 CPU with a clock speed of 2.60 GHz – 2.569 GHz, and 16GB RAM, and running on 64-bit Windows 10 operating system. The execution time for each round is noted in seconds.

TABLE I. CNN Regression Results: Univariate Timeseries with One Week Data as Input (CNN#1)

| No. | RMSE | Mon | Tue | Wed | Thu | Fri | Time |
|---|---|---|---|---|---|---|---|
| 1 | 3.840 | 2.90 | 3.60 | 3.90 | 4.20 | 4.50 | 85.62 |
| 2 | 5.471 | 4.50 | 5.30 | 5.30 | 5.80 | 7.40 | 81.07 |
| 3 | 4.370 | 2.90 | 4.10 | 4.80 | 4.70 | 5.00 | 83.17 |
| 4 | 3.804 | 3.00 | 3.50 | 3.80 | 4.10 | 4.40 | 83.49 |
| 5 | 3.840 | 3.00 | 3.50 | 3.90 | 4.20 | 4.50 | 80.57 |
| 6 | 3.964 | 3.20 | 3.80 | 4.00 | 4.20 | 4.50 | 82.74 |
| 7 | 4.138 | 3.20 | 3.50 | 4.50 | 4.60 | 4.60 | 82.67 |
| 8 | 3.852 | 3.00 | 3.40 | 3.80 | 4.20 | 4.50 | 86.20 |
| 9 | 4.364 | 3.60 | 4.30 | 4.40 | 4.80 | 4.60 | 83.98 |
| 10 | 3.944 | 2.90 | 3.90 | 3.90 | 4.10 | 4.70 | 84.64 |
| Mean | **4.1587** | 3.22 | 3.89 | 4.23 | 4.49 | 4.87 | **83.42** |
| Min | 3.804 | 2.90 | 3.40 | 3.80 | 4.10 | 4.40 | 80.57 |
| Max | 5.471 | 4.50 | 5.30 | 5.30 | 5.80 | 7.40 | 86.20 |
| SD | 0.507 | 0.50 | 0.58 | 0.51 | 0.53 | 0.90 | 1.80 |
| RMSE/Mean | **0.0066** | 0.005 | 0.006 | 0.007 | 0.007 | 0.008 | |

Table I presents the performance results of the CNN#1 model. It is observed that the CNN#1 model takes, on an average, 83.42s for the execution of one round. It yields a mean value of the ratio of RMSE to the mean of the *open* values in the test dataset as 0.0066. The mean RMSE values of the model is found to increase consistently from Monday to Friday. Fig. 6 presents the performance results of the CNN#1 model for round #5 in Table I. The mean RMSEs for the model for Monday to Friday have been found to be 0.005123, 006189, 00673, 0.007144, and 0.007748.

The performance results of the CNN#2 are presented in Table II. The model exhibits a mean execution time of 87.29s for one round, which is higher than that of the model CNN#1 model. The average value of the ratio of the RMSE to the mean of the actual *open* values is 0.0062, which is smaller than the corresponding value of the model CNN#1. The mean RMSE of the model for the five days in a week were 0.004868, 0.005823, 0.006316, 0.006587, and 0.007303 respectively. Fig. 7 depicts the variations of RMSE of the model as per the round #1 in Table II.

The performance results of the CNN#3 model are presented in Table III. While the mean execution time for one round of the model is 116.45s, the RMSE to the mean of the actual *open* values in the test dataset for the model is found to be 0.0090. The mean RMSE values for the different days in a week for the model are found to be 0.008051, 0.008751, 0.008862, 0.009673, and 0.009689 respectively. Fig.8 shows the variations of RMSE with respect to different days in a week as per the round 6 in Table III.

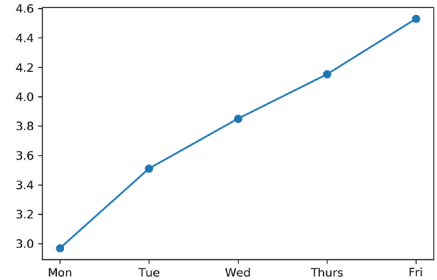

Fig. 6. RMSE of CNN#1 - univariate time series with the previous one-week data as input (Round #5 of Table I)

TABLE II. CNN Regression Results: Univariate Timeseries with Previous Two Weeks Data as Input (CNN#2)

| No. | RMSE | Mon | Tue | Wed | Thu | Fri | Time |
|---|---|---|---|---|---|---|---|
| 1 | 3.994 | 2.70 | 3.90 | 4.10 | 4.20 | 4.80 | 87.80 |
| 2 | 3.544 | 2.50 | 3.20 | 3.50 | 3.90 | 4.30 | 84.92 |
| 3 | 4.833 | 4.60 | 4.30 | 4.60 | 4.80 | 5.70 | 88.05 |
| 4 | 3.774 | 2.80 | 3.50 | 4.00 | 4.10 | 4.20 | 82.78 |
| 5 | 3.913 | 3.10 | 3.70 | 4.10 | 4.10 | 4.40 | 89.22 |
| 6 | 3.739 | 2.50 | 3.30 | 3.90 | 4.00 | 4.60 | 88.48 |
| 7 | 3.583 | 2.70 | 3.30 | 3.60 | 4.00 | 4.20 | 87.23 |
| 8 | 4.454 | 3.70 | 4.40 | 4.50 | 4.10 | 5.40 | 89.82 |
| 9 | 3.635 | 2.50 | 3.40 | 3.60 | 4.30 | 4.10 | 86.38 |
| 10 | 3.796 | 3.50 | 3.60 | 3.80 | 3.90 | 4.20 | 88.26 |
| Mean | **3.9265** | 3.06 | 3.66 | 3.97 | 4.14 | 4.59 | **87.29** |
| Min | 3.544 | 2.50 | 3.20 | 3.50 | 3.90 | 4.10 | 82.78 |
| Max | 4.833 | 4.60 | 4.40 | 4.60 | 4.80 | 5.70 | 89.82 |
| SD | 0.4122 | 0.69 | 0.42 | 0.37 | 0.26 | 0.55 | 2.11 |
| RMSE/Mean | **0.0062** | 0.005 | 0.006 | 0.006 | 0.007 | 0.007 | |

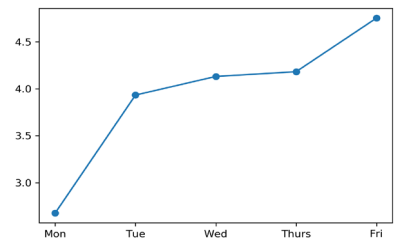

Fig. 7. RMSE of CNN#2 – univariate time series with previous weeks' data as input (Round #1 of Table II)

TABLE III. CNN Regression Results: Multivariate Time Series with Previous Two Weeks' Data as Input (CNN#3)

| No. | RMSE | Mon | Tue | Wed | Thu | Fri | Time |
|---|---|---|---|---|---|---|---|
| 1 | 9.522 | 8.80 | 8.90 | 9.50 | 11.00 | 9.20 | 115.71 |
| 2 | 6.530 | 6.10 | 6.20 | 6.40 | 7.00 | 6.90 | 117.71 |
| 3 | 4.605 | 3.70 | 4.20 | 4.50 | 5.30 | 5.10 | 117.55 |
| 4 | 6.120 | 5.90 | 6.20 | 5.80 | 6.50 | 6.20 | 116.81 |
| 5 | 4.047 | 3.10 | 4.10 | 4.10 | 4.30 | 4.60 | 114.25 |
| 6 | 5.084 | 3.70 | 5.20 | 5.10 | 5.50 | 5.70 | 115.22 |
| 7 | 4.179 | 3.60 | 4.10 | 4.20 | 4.30 | 4.60 | 116.57 |
| 8 | 5.001 | 5.00 | 5.50 | 4.50 | 4.80 | 5.10 | 116.57 |
| 9 | 5.768 | 5.30 | 5.20 | 5.90 | 5.90 | 6.50 | 121.45 |
| 10 | 6.002 | 5.40 | 5.40 | 5.70 | 6.20 | 7.00 | 113.24 |
| Mean | **5.6858** | 5.06 | 5.5 | 5.57 | 6.08 | 6.09 | **116.45** |
| Min | 4.047 | 3.10 | 4.10 | 4.10 | 4.30 | 4.60 | 113.24 |
| Max | 9.522 | 8.80 | 8.90 | 9.50 | 11.00 | 9.20 | 121.45 |
| SD | 1.587 | 1.69 | 1.43 | 1.59 | 1.95 | 1.41 | 2.24 |
| RMSE/Mean | **0.009** | 0.008 | 0.009 | 0.009 | 0.010 | 0.010 | |

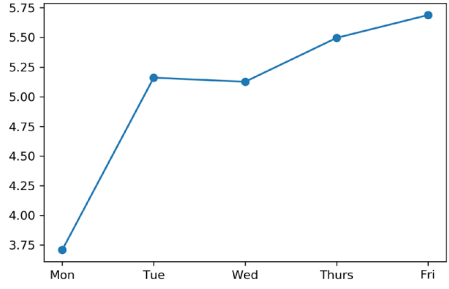

Fig. 8. RMSE of CNN#3 – multivariate time series with the previous two weeks' data as input (Round #6 of Table III)

Table IV presents the performance of the LSTM#1 model. It is observed that the mean time of execution of one round of the model is 330.75s, while the value of the RMSE to the mean *open* value in the test dataset is 0.007. The mean RMSE of the model for different days in a week are 0.006094, 0.006825, 0.00716, 0.007414, and 0.007716 for Monday to Friday respectively. Fig. 9 shows how the mean RMSE of the model varies for different days in a week as per the round 7 in Table IV.

TABLE IV. LSTM REGRESSION RESULTS: UNIVARIATE TIMESERIES WITH PREVIOUS ONE WEEK'S DATA AS INPUT (LSTM#1)

| No. | RMSE | Mon | Tue | Wed | Thu | Fri | Time |
|---|---|---|---|---|---|---|---|
| 1 | 3.967 | 3.30 | 3.50 | 4.10 | 4.20 | 4.60 | 323.55 |
| 2 | 4.773 | 3.30 | 4.60 | 5.50 | 4.80 | 5.10 | 333.50 |
| 3 | 6.976 | 7.60 | 6.50 | 7.20 | 6.80 | 6.70 | 322.84 |
| 4 | 4.466 | 3.00 | 4.60 | 4.20 | 4.80 | 5.40 | 311.59 |
| 5 | 4.408 | 4.00 | 3.60 | 4.80 | 4.80 | 4.70 | 324.14 |
| 6 | 3.569 | 3.10 | 3.20 | 3.40 | 3.90 | 4.10 | 319.44 |
| 7 | 3.807 | 3.00 | 3.70 | 4.10 | 4.00 | 4.10 | 343.46 |
| 8 | 3.510 | 2.60 | 3.00 | 3.50 | 3.80 | 4.40 | 317.20 |
| 9 | 4.297 | 4.60 | 5.30 | 3.30 | 4.10 | 3.90 | 356.06 |
| 10 | 4.928 | 3.80 | 4.90 | 4.90 | 5.40 | 5.50 | 355.71 |
| Mean | **4.4701** | 3.83 | 4.29 | 4.50 | 4.66 | 4.85 | **330.75** |
| Min | 3.510 | 2.60 | 3.00 | 3.30 | 3.80 | 3.90 | 311.59 |
| Max | 6.976 | 7.60 | 6.50 | 7.20 | 6.80 | 6.70 | 356.06 |
| SD | 1.002 | 1.44 | 1.09 | 1.18 | 0.91 | 0.85 | 15.88 |
| RMSE/Mean | **0.007** | 0.0061 | 0.0068 | 0.0072 | 0.0074 | 0.0077 | |

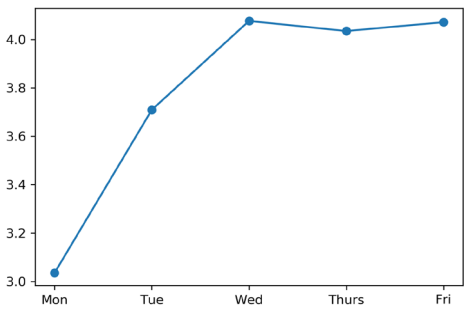

Fig. 9. RMSE of LSTM#1- univariate LSTM time series with the previous one week's data as input (Round #7 of Table IV)

Table V shows that the mean time for execution of one round of the model LSTM#2 is 544.2s, while the value of the ratio of RMSE to the mean of the actual *open* values in the test dataset is 0.007. The mean daily RMSE of the model for different days in a week are 0.006253, 0.006253, 0.007223, 0.007573, and 0.007828 respectively. Fig. 10 depicts the way the mean RMSE of the model varies with different days in a week as per the round 10 in Table V.

TABLE V. LSTM REGRESSION RESULTS: UNIVARIATE TIMESERIES WITH PREVIOUS TWO WEEKS' DATA AS INPUT (LSTM#2)

| No. | RMSE | Mon | Tue | Wed | Thu | Fri | Time |
|---|---|---|---|---|---|---|---|
| 1 | 9.385 | 11.00 | 9.00 | 8.90 | 10.00 | 7.70 | 546.27 |
| 2 | 4.562 | 3.30 | 3.90 | 4.20 | 5.80 | 5.20 | 545.50 |
| 3 | 5.194 | 5.30 | 4.40 | 6.50 | 4.50 | 5.00 | 544.37 |
| 4 | 3.318 | 2.20 | 2.80 | 3.70 | 3.60 | 4.00 | 551.31 |
| 5 | 3.889 | 3.00 | 3.50 | 4.10 | 4.00 | 4.60 | 536.71 |
| 6 | 3.477 | 2.20 | 2.90 | 3.50 | 4.00 | 4.40 | 539.68 |
| 7 | 4.098 | 3.00 | 3.00 | 4.30 | 4.30 | 5.50 | 543.20 |
| 8 | 3.790 | 3.40 | 3.90 | 3.40 | 3.70 | 4.50 | 549.64 |
| 9 | 3.294 | 2.60 | 2.80 | 3.20 | 3.70 | 4.00 | 545.38 |
| 10 | 3.681 | 3.30 | 3.10 | 3.60 | 4.00 | 4.30 | 542.11 |
| Mean | **4.469** | 3.93 | 3.93 | 4.54 | 4.76 | 4.92 | **544.42** |
| Min | 3.294 | 2.20 | 2.80 | 3.20 | 3.60 | 4.00 | 536.71 |
| Max | 9.385 | 11.00 | 9.00 | 8.90 | 10.00 | 7.70 | 551.31 |
| SD | 1.824 | 2.63 | 1.86 | 1.79 | 1.95 | 1.09 | 4.33 |
| RMSE/Mean | **0.007** | 0.006 | 0.006 | 0.007 | 0.008 | 0.008 | |

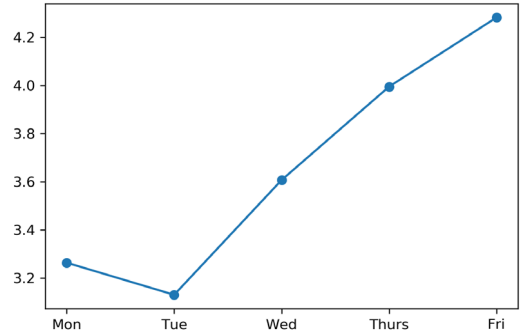

Fig. 10. RMSE of LSTM#2- univariate LSTM model with the previous two weeks' data as input (Round #10 of Table V)

TABLE VI. ENCODER-DECODER LSTM REGRESSION RESULTS: UNIVARIATE TIMESERIES WITH TWO WEEKS DATA AS INPUT (LSTM#3)

| No. | RMSE | Mon | Tue | Wed | Thu | Fri | Time |
|---|---|---|---|---|---|---|---|
| 1 | 6.315 | 6.10 | 6.40 | 6.50 | 6.30 | 6.20 | 307.80 |
| 2 | 7.979 | 7.90 | 7.50 | 7.70 | 7.70 | 9.00 | 299.85 |
| 3 | 3.710 | 2.30 | 3.50 | 3.90 | 4.10 | 4.40 | 321.85 |
| 4 | 5.262 | 5.40 | 5.20 | 5.10 | 5.30 | 320.12 | |
| 5 | 4.891 | 4.90 | 5.10 | 4.90 | 4.80 | 4.70 | 282.49 |
| 6 | 3.489 | 2.50 | 3.10 | 3.50 | 3.90 | 4.20 | 279.04 |
| 7 | 5.232 | 5.00 | 5.40 | 5.50 | 4.80 | 5.40 | 308.66 |
| 8 | 4.102 | 3.50 | 4.10 | 4.10 | 4.30 | 4.50 | 314.91 |
| 9 | 4.024 | 2.90 | 3.60 | 4.10 | 4.50 | 4.70 | 314.40 |
| 10 | 7.747 | 5.90 | 7.60 | 8.00 | 8.40 | 8.60 | 315.02 |
| Mean | **5.275** | 4.64 | 5.16 | 5.34 | 5.39 | 5.70 | **306.41** |
| Min | 3.489 | 2.30 | 3.10 | 3.50 | 3.90 | 4.20 | 279.04 |
| Max | 7.979 | 7.90 | 7.60 | 8.00 | 8.40 | 9.00 | 321.85 |
| SD | 1.607 | 1.81 | 1.62 | 1.59 | 1.56 | 1.74 | 14.93 |
| RMSE/Mean | **0.0083** | 0.007 | 0.008 | 0.008 | 0.009 | 0.009 | |

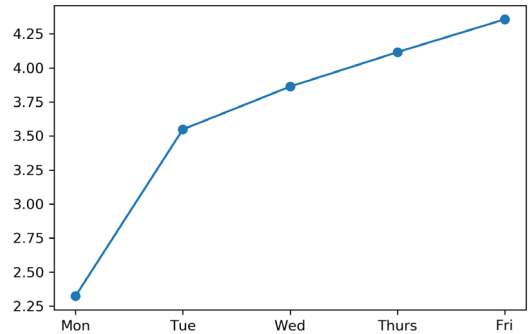

Fig. 11. RMSE of LSTM#3 - univariate encoder-decoder LSTM model with the previous two weeks' data as input (Round #3 of Table VI)

TABLE VII. ENCODER-DECODER LSTM REGRESSION RESULTS: MULTIVARIATE TIMESERIES WITH TWO WEEKS INPUT (LSTM#4)

| No. | RMSE | Mon | Tue | Wed | Thu | Fri | Time |
|---|---|---|---|---|---|---|---|
| 1 | 8.264 | 8.10 | 8.00 | 8.30 | 8.40 | 8.50 | 831.07 |
| 2 | 9.751 | 9.50 | 9.90 | 9.60 | 9.90 | 9.90 | 784.41 |
| 3 | 6.038 | 5.50 | 5.90 | 6.10 | 6.20 | 6.40 | 844.70 |
| 4 | 5.620 | 5.10 | 5.40 | 5.70 | 5.80 | 6.00 | 853.61 |
| 5 | 4.935 | 3.50 | 4.20 | 4.50 | 4.70 | 5.00 | 853.15 |
| 6 | 10.447 | 10.20 | 10.30 | 10.50 | 10.50 | 10.70 | 825.15 |
| 7 | 4.882 | 4.40 | 4.60 | 4.90 | 5.10 | 5.30 | 860.83 |
| 8 | 10.320 | 10.30 | 10.10 | 10.30 | 10.40 | 10.60 | 849.05 |
| 9 | 7.859 | 7.40 | 7.80 | 7.90 | 8.00 | 8.20 | 832.85 |
| 10 | 4.464 | 3.70 | 4.20 | 4.60 | 4.80 | 5.00 | 854.03 |
| Mean | **7.204** | 6.77 | 7.04 | 7.24 | 7.38 | 7.56 | **838.92** |
| Min | 4.395 | 3.50 | 4.20 | 4.50 | 4.70 | 5.00 | 784.41 |
| Max | 10.447 | 10.30 | 10.30 | 10.50 | 10.50 | 10.70 | 860.83 |
| SD | 2.425 | 2.67 | 2.49 | 2.37 | 2.35 | 2.31 | 22.39 |
| RMSE/Mean | **0.0115** | 0.011 | 0.011 | 0.012 | 0.012 | 0.012 | |

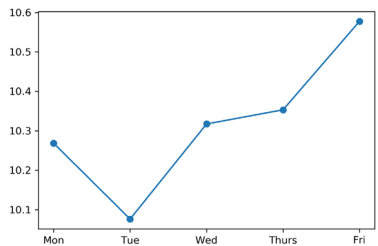

Fig. 12. RMSE of LSTM#4 - multivariate encoder-decoder LSTM model with the previous two weeks' data as input (Round #8 of Table VII)

The performance results of the model LSTM#3 are presented in Table VI. The model takes a mean time of 306.41s for each round, and it yields a value of 0.0083 as the ratio of the RMSE to the mean *open* value. The mean RMSE for different days in a week were found to be 0.007382, 0.00821, 0.008496, 0.008576, and 0.009069 from Monday to Friday respectively. Fig.11 shows the variations of RMSE of the model as per the round 3 in Table VI.

Table VII presents the performance of the LSTM#4 model. The model needs 838.92s, on an average, for each round. The ratio of the RMSE to the mean *open* value in the test dataset yielded by the model is 0.0115. The mean RMSE values for the model with respect to different days in a week are 0.010771, 0.011201, 0.011519, 0.011742, and 0.012028 from Monday to Friday respectively. Fig. 12 shows how the RMSE varies over different days in a week for LSTM#4 as per the round 8 in Table VII.

TABLE VIII. COMPARISON OF DIFFERENT MODELS BASED ON EXECUTION TIME AND RMSE

| Execution Time | | | RMSE/Mean | | |
|---|---|---|---|---|---|
| Rank | Model | Value | Rank | Model | Value |
| 1 | CNN #1 | 83.42s | 1 | CNN #2 | 0.00625 |
| 2 | CNN #2 | 87.29s | 2 | CNN #1 | 0.00662 |
| 3 | CNN #3 | 116.45s | 3 | LSTM #2 | 0.00710 |
| 4 | LSTM #3 | 306.41s | 4 | LSTM #1 | 0.00711 |
| 5 | LSTM #1 | 330.75s | 5 | LSTM #3 | 0.00839 |
| 6 | LSTM #2 | 544.42s | 6 | CNN #3 | 0.00905 |
| 7 | LSTM #4 | 838.92s | 7 | LSTM #4 | 0.11461 |

Table VIII provides a comparative analysis of all the seven models that we have presented in this work. The models are ranked based on two metrics: *execution time*, and *the ratio of RMSE to the mean open value in the test dataset*. It is observed that while the CNN #1 model is found to be the fastest in execution, the model CNN #2 is the most accurate. Overall, the CNN models are found to have outperformed their LSTM counterparts on both the metrics.

## V. CONCLUSION

In this paper, we have presented a suite of the deep learning-based regression model for forecasting stock price on a daily basis for a forecast horizon of one week. While three models are built on CNN architecture, four regression models are designed following LSTM networks. The models were constructed using optimized hyperparameters and then tested on extremely granular stock price data collected at an interval of five minutes. Experimental results showed that while the models exhibited wide divergence in their accuracy and execution speeds, all of them yielded a very high level of accuracy in their forecasting results. In general, the CNN-based models outperformed their LSTM counterparts both in terms of their execution speed and accuracy in prediction.